\documentclass[aip,jcp,amsmath,amssymb,reprint]{revtex4-1}

\usepackage{graphicx}
\usepackage{dcolumn}
\usepackage{bm}
\usepackage{color}

\usepackage[utf8]{inputenc}
\usepackage[T1]{fontenc}
\usepackage{mathptmx}

\newcommand{\hdf}{H_{\rm DF}}
\newcommand{\fdf}{F_{\rm DF}}
\newcommand{\wdf}{W_{\rm DF}}
\newcommand{\tdf}{T_{\rm DF}}
\newcommand{\edf}{E_{\rm DF}}

\newcommand{\tonedf}{T_1^{\rm DF}}
\newcommand{\ttwodf}{T_2^{\rm DF}}

\begin{document}

\preprint{AIP/123-QED}

\title[Reduction of density-fitting error in coupled-cluster calculations]{A
straightforward \emph{a posteriori} method for reduction of density-fitting error in coupled-cluster 
calculations}

\author{Micha\l\ Lesiuk}
\email{lesiuk@tiger.chem.uw.edu.pl}
\affiliation{\sl Faculty of Chemistry, University of Warsaw, Pasteura 1, 02-093 Warsaw, Poland}

\date{\today}

\begin{abstract}
We present a simple method for \emph{a posteriori} removal of a significant fraction of the 
density-fitting error from the calculated total coupled-cluster energies. The method treats the 
difference between the exact and density-fitted integrals as a perturbation, and simplified
response-like equations allow to calculate improved amplitudes and the corresponding energy 
correction. The proposed method is tested at the coupled-cluster singles and doubles level of 
theory for a diverse set of moderately-sized molecules. On average, error reductions by a factor of 
approximately 
ten and twenty are observed in double-zeta and triple-zeta basis sets, respectively. Similar 
reductions are observed in calculations of interaction energies of several model complexes. The 
computational 
cost of the procedure is small in comparison with the preceding coupled-cluster iterations. The 
applicability of the method is not limited to the density-fitting approximation; in principle, it 
can be used in conjunction with an arbitrary decomposition scheme of the electron repulsion 
integrals.
\end{abstract}

\maketitle

\section{\label{sec:intro} Introduction}

Density fitting (DF) approximation\cite{whitten73,baerends73,dunlap79,alsenoy88,vahtras93} is a 
popular method used to reduce the computational burden and 
storage requirements related to handling of electron repulsion integrals (ERIs) present in most 
quantum chemistry methods. It relies on the following formula
\begin{align}
 \label{dfint}
 (\mu\nu|\lambda\sigma) = (\mu\nu|P)\,[\mathbf{V}^{-1}]_{PQ}(Q|\lambda\sigma),
\end{align}
where the Greek letters denote the atomic (orbital) basis set, $(\mu\nu|P)$ and $V_{PQ}=(P|Q)$ are 
the three-centre and two-centre electron repulsion integrals, 
respectively (see Ref.~\onlinecite{katouda09} for details of the notation). The the $P$, $Q$ 
summation runs over 
elements of the auxiliary basis set (ABS) that is designed to balance the accuracy of the 
method against its computational cost. In most modern applications ABS are pre-optimized for 
a given orbital basis set family and a level of theory. When only ERIs are concerned the DF 
approach is equivalent to (a more general) resolution-of-identity approximation, and thus in the 
present context both names are used interchangeably in the literature.

The DF approximation has originally been proposed to simplify the self-consistent field 
calculations, but its use has been extended to various other electronic structure 
methods, including sophisticated explicitly correlated\cite{manby03,werner07} and coupled-cluster 
theories\cite{schutz03,kats07,korona08,bostrom12,deprince13,epifanovsky13,deprince14}. Much effort 
has also been put into improving the basic approximation 
formula\cite{wiegend09,grajciar15,manzer15,duchemin17}, exploiting sparsity in the three-center 
integrals \cite{hohenstein12a,parrish12,hohenstein12b} and design more straightforward and/or 
automated ways to generate ABS\cite{weigend02,jensen05,hattig05,stoychev17}.

The DF approximation has been shown to deliver accuracies better than 0.1 kJ/mol in relative 
energies\cite{deprince13}, even if ABS has been optimized with a different level of theory in mind. 
However, this 
success relies solely on error cancellation. The error in the total energies can, in fact, be 
substantial and there are several reasons that motivate the development of methods for its 
reduction. For example, ABS are most commonly optimized and used in calculations for molecules near 
their equilibrium geometric structures. This does not guarantee that the errors are consistently 
small, e.g., for molecular complexes away from their equilibrium geometries.

An \emph{a posteriori} procedure for elimination of DF error from the total energies 
has been recently proposed by Schurkus et al\cite{schurkus17}. It starts with an oversized 
auxiliary basis set that 
allows to eliminate a great portion of the error, and then projects out linear combinations of 
functions that do not contribute significantly at a given level of theory. Here we present an 
alternative approach that is designed to be used in conjunction with the coupled-cluster method. It is rooted in 
perturbation theory, with the error in the integrals treated as a perturbation, and allows to 
derive response-like equations for the perturbed amplitudes.

While in this work we concentrate on the DF approximation, the proposed approach is 
applicable to arbitrary ERIs decomposition schemes. Therefore, it can be combined with, e.g., the 
Cholesky decomposition\cite{beebe77,koch03,pedersen04,folkestad19} or the pseudospectral 
method\cite{friesner86,martinez94}, or even the recently proposed tensor 
hypercontraction\cite{hohenstein12a,parrish12,hohenstein12b,parrish13}, canonical product 
format\cite{benedikt11,benedikt13} or chain-of-spheres 
algorithm\cite{neese09,kossmann10,izsak11,taras11,izsak12,izsak13,dutta16}.

\section{\label{sec:theory} Theory}

The coupled cluster theory\cite{crawford07,bartlett07} relies on exponential representation of the 
electronic wavefunction, 
$|\Psi\rangle = e^T\,|\phi_0\rangle$. In this work we consider coupled-cluster singles and doubles 
(CCSD) model\cite{purvis82,scuseria87}, i.e., $T=T_1+T_2$ with
\begin{align}
\label{cluster}
 T_1 = \sum_{ai} t_i^a \,a^\dagger i,\;\;\;
 T_2 = \frac{1}{4}\sum_{abji} t_{ij}^{ab} \,a^\dagger b^\dagger j i,
\end{align}
where the creation $a^\dagger, b^\dagger,\ldots$, and annihilation operators $i,j,\ldots$ are 
defined as in Ref.~\onlinecite{bartlett07}. The cluster amplitudes ($t_i^a, 
t_{ij}^{ab}$) are obtained by solving the 
equations
\begin{align}
\label{projn}
 \langle\mu_n|\,e^{-T} H e^T\rangle = 0,
\end{align}
where $H$ is the electronic Hamiltonian, and $\langle\mu_n|$ denotes projection on the manifold of 
singly ($n=1$) or doubly ($n=2$) excited state determinants. The electronic energy is calculated 
as $E = \langle e^{-T} H e^T\rangle$. 

For the purposes of this work we introduce the second Hamiltonian operator, 
$\hdf$. It has exactly the same second-quantized form as $H$, but the usual 
two-electron integrals are replaced by their DF counterparts. While a basis set that exactly spans 
the finite orbital product space exists\cite{pedersen09} (making the relationship $\hdf=H$ strictly 
valid), it is not true for the optimized atom-centered basis sets that are used in practice, and 
thus the relevant equations of the DF-CCSD method read
\begin{align}
 \langle\mu_n|\,e^{-\tdf} \hdf \, e^{\tdf}\rangle = 0, \;\;\;
 \edf = \langle e^{-\tdf} \hdf\, e^{\tdf}\rangle,
\end{align}
with $\tdf\neq T$. To introduce this formalism we assumed that the DF approximation is applied only 
at the correlated level, i.e., the Hartree-Fock orbitals and the corresponding 
creation/annihilation operators are the same in $T$ and $\tdf$. 

It is also useful to divide the 
Hamiltonian into a sum of the Fock operator and the fluctuation potential, $H=F+W$, and similarly 
for the DF counterpart, $\hdf=\fdf+\wdf$. Note that the despite the same orbitals are 
used to define $F$ and $\fdf$ one still has $F\neq\fdf$ since the Fock operator itself depends on 
the two-electron integrals. This results in a non-vanishing occupied-virtual block of the $\fdf$ 
matrix and a violation of the Brillouin condition. However, as pointed out earlier in the 
literature\cite{deprince13}, this formal problem typically has little effect on the calculated 
correlation energies. 
For this reason we set $F=\fdf$ throughout the present work. This not only simplifies the formalism 
but also makes the final working equations straightforward to solve.

The main goal of this paper is to define a perturbative correction that approximates the error in 
the correlation energy due to the DF approximation
\begin{align}
 \delta E = E - \edf,
\end{align}
and can simply be added on top of the converged DF-CCSD energy. To this end we define the amplitudes 
correction $\delta T=T-\tdf$, and attempt to find an expression for $\delta T$ that is linear in 
$\delta W=H-\hdf=W-\wdf$. Next we insert these two expressions into Eq. (\ref{projn}) and eliminate 
$H$ and $T$ from the final formula
\begin{align}
\label{proj2}
\begin{split}
 0 &= \langle\mu_n|\,e^{-\tdf-\delta T} \hdf\, e^{\tdf+\delta T}\rangle \\
   &+ \langle\mu_n|\,e^{-\tdf-\delta T} \delta W\, e^{\tdf+\delta T}\rangle.
\end{split}
\end{align}
By applying the nested commutator expansion (Baker-Campbell-Hausdorff expansion) one arrives at
\begin{align}
\begin{split}
 0 &= \langle\mu_n|\,\Big[e^{-\tdf} \hdf\, e^{\tdf},\delta T\Big]\rangle \\
   &+ \langle\mu_n|\,e^{-\tdf} \delta W\, e^{\tdf}\rangle + \mathcal{O}\left(\delta^2\right),
\end{split}
\end{align}
where the notation $\mathcal{O}\left(\delta^2\right)$ signals that the remaining terms are at least 
quadratic in the combined powers of $\delta W$ and $\delta T$. The zeroth-order term vanishes since 
$\langle\mu_n|\,e^{-\tdf} \hdf e^{\tdf}\rangle$ is the DF-CCSD stationary condition. With all 
higher-order terms neglected the resulting equation can be brought into a more familiar form
\begin{align}
\label{resp}
 \langle\mu_n|\,\Big[e^{-\tdf} \hdf\, e^{\tdf},\delta T\Big]\rangle
 = -\langle\mu_n|\,e^{-\tdf} \delta W\, e^{\tdf}\rangle,
\end{align}
This equation is linear in $\delta T$ and is closely related to the CCSD linear 
response amplitude equations\cite{monkhorst77,dalgaard83,koch90,kobayashi94}; the matrix on the left-hand side 
is the so-called coupled-cluster Jacobian. From 
this point of view, our method can be understood as variant of the usual response theory applied to 
the perturbation in the two-electron integrals, $\delta W$. By using similar arguments one can find the desired energy 
correction
\begin{align}
\label{ecorr}
\begin{split}
 \delta E &= \langle\Big[\wdf,\,\delta T_2+\tonedf\delta T_1\Big]\rangle \\
  &+ \langle \Big[\delta W,\,\ttwodf+{\textstyle {\frac12}}\left(\tonedf\right)^2\Big]\rangle + 
\mathcal{O}\left(\delta^2\right),
\end{split}
\end{align}
where the partitions $\tdf=\tonedf+\ttwodf$ and $\delta T=\delta T_1+\delta T_2$ have been 
introduced.

Unfortunately, Eq. (\ref{resp}) and (\ref{ecorr}) that constitute the basis of our formalism 
would not be computationally beneficial at this point. Solution of the response equations 
(\ref{resp}) requires virtually as much computational effort per iteration as the DF-CCSD method. 
The only gain one could hope for is that Eq. (\ref{resp}), being linear in $\delta T$, may 
require less iterations to converge than the (inherently non-linear) coupled-cluster equations. All 
in all, it would still probably be more cost-effective to simply run DF-CCSD calculations with a 
larger auxiliary basis set.

To circumvent this difficulty we proceed with expansion of $\delta T$ in orders of 
the unperturbed fluctuation potential, $\wdf$. In other words, we write
\begin{align}
\label{dtpert}
 \delta T_1 = \delta T_1^{(0)} + \delta T_1^{(1)} + \ldots,
\end{align}
and analogously for $\delta T_2$, where the superscripts denote the order in $\wdf$. We treat $F$, 
$\delta W$, and $\wdf$ as zeroth-, zeroth-, and first-order quantities, respectively. Under these 
assumptions the cluster operators $\ttwodf$ and $\tonedf$ enter in the first and second order, 
respectively. Order-by-order expressions for the perturbed cluster operators $\delta T_n^{(m)}$ are 
obtained by expanding the exponentials in Eq. (\ref{resp}) with help of the nested commutator 
formula, inserting Eq. (\ref{dtpert}), and collecting terms of the same order. Here it sufficient 
to consider only the zeroth and first order corrections that are defined by the equations
\begin{align}
\langle \mu_2|\Big[F,\delta T_2^{(0)}\Big]+\delta W\rangle=0,
\end{align}
and
\begin{align}
\label{t21}
&\langle \mu_n|\Big[F,\delta T_n^{(1)}\Big]+\Big[\wdf, \delta T_2^{(0)}\Big]+\Big[\delta W, 
\ttwodf\Big] \rangle=0,
\end{align}
where $n=1,2$, and $T_1^{(0)}=0$. The advantage of the above formulas in comparison with the 
initial Eq. (\ref{resp}) is that there is no need to solve linear equations for $\delta T_n^{(m)}$ 
-- the operator $F$ is diagonal is the chosen basis and the equations are inverted in a one-step 
procedure. 

The dominant portion of the computational cost related to Eq. (\ref{t21}) is due to contributions 
involving $\delta W$ since they require evaluation of the exact two-electron integrals. In our 
pilot implementation we followed the direct CCSD approach of Koch and 
collaborators\cite{koch94,koch96} to treat these 
terms, but exploited various simplifications resulting from the linear nature of the above 
equations. Computation of $\delta T_n^{(m)}$ and the energy correction (\ref{ecorr}) is thus less 
expensive than a single iteration of the conventional CCSD method, mostly due to absence of 
all terms quadratic in the amplitudes. Moreover, it can be accomplished without storage of any 
intermediate quantities on the disk and the associated I/O costs. Overall, we believe that the cost 
of evaluating Eqs. (\ref{ecorr}) and (\ref{t21}) is a reasonable price to pay for a sizable 
reduction of the DF error observed in benchmark calculations reported in the next 
section.

To finalize this section let us point out that all working equations of the proposed method are 
expressed solely in terms of commutators of connected quantities. Therefore, the resulting theory 
contains no disconnected terms and thus is rigorously size extensive. Moreover, the correction 
$\delta E$ given by Eqs. (\ref{ecorr}) and (\ref{t21}) vanishes identically in the limit of 
complete auxiliary basis set, i.e., when $\delta W\rightarrow 0$.

\section{\label{sec:num} Numerical results}

\subsection{Computational details}

The perturbative method described in the previous section was implemented in a locally modified 
version of the \textsc{Gamess} program package\cite{schmidt93} employing a set of DF-CC codes 
written by the present author. The code handling the DF decomposition of the two-electron integrals relies on the 
resolution-of-identity MP2 (RI-MP2) implementation by Katouda and Nagase\cite{katouda09}. 
Hartree-Fock equations 
were solved without any approximations to the two-electron integrals.

In this study the Dunning-family basis sets\cite{dunning89,kendall92,woon93} cc-pVDZ and cc-pVTZ, 
along with their augmented (aug-) counterparts, were employed in the calculations. For the 
auxiliary basis expansion the accompanying 
MP2FIT basis sets optimized by Weigend et al.\cite{weigend98,weigend02} were used. Pure spherical 
representations of both the 
orbital and the auxiliary basis sets were used throughout.

\subsection{Total energies}

For testing the performance of the proposed approach for the total correlation energies the diverse 
benchmark set developed by Adler and Werner was chosen\cite{adler11}. It consists of 71 molecular 
systems ranging 
in size from two up to eighteen atoms, and from two up to about fifty active electrons. The 
geometries of the molecules can be found in Ref. \onlinecite{adler11}. For the first row atoms the 
$1s$ core orbitals 
were frozen (inactive) in the correlated calculations and for the second row atoms the same applied 
to the $2s$ and $2p$ orbitals. 

For each molecule in the test test we performed separate DF-CCSD and CCSD calculations, and 
computed the $\delta E$ correction according to the prescription given in the previous section. 
Next we calculated the relative error in the obtained correlation energies with and without 
application of the correction term, that is
\begin{align}
\label{err0}
 10^3 \left| \frac{E_{\rm DF-CCSD} - E_{\rm CCSD}}{E_{\rm CCSD}} \right|,
\end{align}
and
\begin{align}
\label{err1}
 10^3 \left| \frac{E_{\rm DF-CCSD} + \delta E - E_{\rm CCSD}}{E_{\rm CCSD}} \right|,
\end{align}
where $E_{\rm DF-CCSD}$ and $E_{\rm CCSD}$ are the raw DF-CCSD and CCSD correlation energies, 
respectively. For better readability the results are given in parts per thousand, i.e., multiplied 
by a factor of $10^3$.

Since the number of molecules included in the test set is substantial we can perform statistical 
analysis of the errors to gauge the performance of our method without a bias. We thus calculated 
the mean and median relative errors over the whole test set, see Table \ref{tab:data}. It is also 
typical to report the standard deviation in such analysis, but in our case it was very susceptible 
to the outliers with (accidentally) small errors. Therefore, in Table \ref{tab:data} we 
instead report the median deviation which does not suffer from this problem, along with the maximum 
and minimum relative error found in the test set.

\renewcommand{\arraystretch}{1.2}
\begin{table}[t!]
\caption{\label{tab:data}
Statistical measures of the relative density-fitting error (in parts per thousand) in the DF-CCSD 
correlation energy  with and without the $\delta E$ correction term; see Eqs. (\ref{err0}) and 
(\ref{err1}) for precise definitions. The statistics comes from 71 molecules contained in the 
Adler-Werner benchmark set\cite{adler11}.
}
\begin{ruledtabular}
\begin{tabular}{lcc}
 error measure & uncorrected & corrected \\
\hline
\multicolumn{3}{c}{cc-pVDZ} \\
\hline
mean        & 0.714 & 0.074 \\
median      & 0.563 & 0.052 \\
median dev. & 0.100 & 0.022 \\
max.        & 3.089 & 0.563 \\
min.        & 0.361 & 0.009 \\
\hline
\multicolumn{3}{c}{cc-pVTZ} \\
\hline
mean        & 0.726 & 0.042 \\
median      & 0.626 & 0.035 \\
median dev. & 0.023 & 0.011 \\
max.        & 2.007 & 0.201 \\
min.        & 0.577 & 0.006 \\
\end{tabular}
\end{ruledtabular}
\end{table}

From Table \ref{tab:data} one can see that the performance of the perturbative approach depends 
significantly on the quality of the orbital basis set. In the smaller cc-pVDZ basis the DF error is reduced by a factor 
of about ten, both in the mean and in the median. This 
is accompanied by approximately fivefold reduction of the median deviation and the maximum 
deviation. In the larger cc-pVTZ basis set these gains are even larger -- the mean and median 
errors are reduced approximately twenty times. It may be surprising at first that the perturbative 
approach is more successful in the triple-zeta basis set. This may be due to the fact that the 
number of auxiliary functions increases with the size of the orbital basis set and 
thus the operator $\delta W$ constitutes a ``smaller'' perturbation, justifying the neglect of 
higher-order terms in the derivations presented in Section \ref{sec:theory}. Based on 
experience gained from the calculations for the Adler-Werner benchmark set we also note that the 
terms in Eq. (\ref{ecorr}) that contain the $\delta T_1$ operator bring a very small contribution 
to $\delta E$, typically of the order of tens of nanohartrees. It is thus reasonable to neglect 
these terms in future implementations, simplifying the present formalism even further.

At this point we can also compare the performance of the perturbative approach with 
other methods one may employ to reduce the DF error. The simplest solution is to increase the size 
of the auxiliary basis set by one cardinal number, e.g., use orbital basis set cc-pVDZ combined 
with cc-pVTZ-RI auxiliary basis. This approach is especially attractive due to its simplicity, 
provided that a suitable DF basis is available in the literature, and lack of changes in the 
implementation. To investigate the performance of this approach we repeated the calculations for 
the whole Adler/Werner benchmark set, but using cc-pVTZ-RI auxiliary basis in combination with the 
cc-pVDZ orbital basis and cc-pVQZ-RI with the cc-pVTZ orbital basis. Increasing the size 
of auxiliary basis set by one cardinal number reduces the error in the DF-CCSD energies by about 
one order of magnitude, both in cc-pVDZ and cc-pVTZ bases. For example, the average 
relative error in the DF-CCSD/cc-pVDZ energies becomes 0.052 (parts per thousand), compared with 
0.714 for the uncorrected results, and 0.074 for the perturbative approach, in the same units. 
Therefore, the method of increasing the basis set has a slight advantage in the cc-pVDZ basis set, 
and this trend is reversed in the cc-pVTZ basis (cf. Table \ref{tab:data}). We also checked that 
these conclusions remain virtually unaffected in calculation of relative energies reported in the 
next section.

It is also mandatory to 
compare both methods in terms of their computational timings. One can argue that increasing the 
size of the auxiliary basis set should have an advantage in this respect because the 
rate-determining step in the CCSD iterations scales as $O^2V^4$, where $O$ is the number of occupied 
orbitals and $V$ is the number of virtual orbitals. Therefore, the cost of this step in unaffected 
by the size of the DF basis, denoted by $N_{\rm DF}$ further in the text, and determination of the 
density-fitted two-electron integrals introduces only an overhead proportional to $N_{\rm DF} V^4$. 
To compare the timings we consider the nitrobenzole molecule in the cc-pVDZ orbital basis set which 
constitutes the worst-case scenario for the perturbative approach in Adler/Werner benchmark set 
($O=32$, $N_{\rm DF}=573$ for cc-pVDZ-RI and $N_{\rm DF}=879$ for cc-pVTZ-RI). The overhead caused 
by increase of the size of the auxiliary basis set to cc-pVTZ-RI is 46\% of the total DF-CCSD 
iterations time in relation to the calculations with the cc-pVDZ-RI basis. By comparison, 
calculation of the perturbative correction incurs only about 5\% penalty in relation to the 
DF-CCSD/cc-pVDZ timings. While this gap is expected to decrease with the number of electrons in the 
system, the proposed method still appears to be competitive for a broad range of molecules, 
especially taking into consideration the aforementioned absence of I/O costs.

Another idea one may employ to reduce the DF error is
to take the converged DF-CCSD amplitudes and evaluate the coupled-cluster energy by using the exact 
integrals. This approach corresponds to neglecting the first term in Eq. (\ref{ecorr}) responsible 
for ``relaxation'' of the amplitudes. However, in our test calculations we found that this term is 
actually dominant and its omission typically leads to a correction that is by an order 
of magnitude too small and, in many cases, of a wrong sign. The second idea would be to converge 
the DF-CCSD iterations to a certain threshold and then perform a single iteration of the 
conventional CCSD (using the standard Jacobi update of the amplitudes). We tested this approach for 
several molecules from the Adler-Werner test set and found that it is able to reduce the error, on 
average, by a factor of $2-3$. However, to reach the accuracy levels comparable to the perturbative 
method a larger number of iterations is required. Another problem of this approach is that the 
amplitudes obtained in this way may violate the size-consistency requirement and do not fulfill any 
particular stationary condition (to a reasonable accuracy). Therefore, it is not obvious how to, 
e.g., evaluate the molecular properties, calculate corrections accounting for the higher 
excitations, etc.

\subsection{Relative energies}

\renewcommand{\arraystretch}{1.2}
\begin{table}[t!]
\caption{\label{tab:datarel}
Density-fitting error in the DF-CCSD(T) interaction energy (in cm$^{-1}$) with and without the 
$\delta E$ correction term (see text for details of the computations). Total CCSD(T) 
interaction energies (in cm$^{-1}$) for the complexes are given in the second column.}
\begin{ruledtabular}
\begin{tabular}{lrcc}
 system & \multicolumn{1}{c}{total} & uncorrected & corrected \\
\hline
water dimer              & 1665.6 & 0.486 & 0.083 \\
NH$_3\cdots$CH$_4$       & 252.1  & 0.207 & 0.010 \\
methane dimer            & 174.2  & 0.115 & 0.004 \\
\hline
benzene-methane          & 98.2   & 1.180 & 0.163 \\
benzene dimer, stack     & 362.4  & 1.352 & 0.764 \\
benzene dimer, T-shaped  & 56.3   & 4.571 & 0.076 \\
pyrazine dimer           & 358.2  & 4.234 & 0.165 \\
pyridoxine-aminopyridine & 4576.5 & 2.210 & 1.090 \\
\end{tabular}
\end{ruledtabular}
\end{table}

While it has been shown that the perturbative approach is successful in reducing the 
DF error in the total correlation energies, it is interesting to check how this 
translates 
into accuracy of relative energies. For this purpose we evaluated interaction energies for 
three complexes from the A24 data set of \v{R}ez\'{a}\u{c} and Hobza\cite{rezac13}. We selected 
systems with different bonding characters: water dimer (hydrogen bond), ammonia-methane complex 
(mixed induction/dispersion) and methane dimer (dispersion). Calculations were performed at the 
CCSD(T)/aug-cc-pVTZ level of theory\cite{ragha89} with frozen $1s$ core orbitals. The perturbative 
correction was applied to the raw CCSD energies without any change to the (T) component of the 
interaction energy. The latter term was calculated according to the algorithm described in 
Ref.~\onlinecite{deprince13}. To test the performance of the proposed method for interaction 
energies between larger molecules we additionally provide results for five systems from the 
S22 benchmark set\cite{jurecka06}. These are: benzene-methane complex, benzene dimer (parallel 
displaced and T-shaped geometries), pyrazine dimer, and pyridoxine-aminopyridine complex. The 
CCSD(T)/cc-pVDZ level of theory was applied for the larger systems, also within the frozen-core 
approximation.

The results are reported in Table \ref{tab:datarel}. The perturbative correction reduces the error 
in the relative energies by a factor comparable to the total energies. This means that after adding 
the $\delta E$ term the discrepancies in the interaction energies are reduced to a level below
$1$ cm$^{-1}$. We believe that this accuracy would be sufficient for a majority 
applications, even aiming at spectroscopic applications, and would make the DF-CCSD(T) and 
conventional CCSD(T) results indistinguishable in such cases. Note that the proposed method appears to be particularly 
advantageous for dispersion-bound complexes where the interaction energies are positive only at correlated levels 
of theory and thus more affected by the density-fitting error on a relative basis.

\section{\label{sec:concl} Conclusions}

We have presented a systematic approach to derive perturbative \emph{a posteriori} corrections that 
remove the bulk of error due to DF approximation from the calculated coupled-cluster 
energies. The simplest variant of the method has been tested for total and relative energies of 
numerous benchmark systems revealing a systematic improvement in the quality of the results. 
An additional advantage of the method is that it can be applied to other decompositions of 
the two-electron integrals, besides the DF approximation considered in this work. In 
future we plan to extend the formalism reported here to more accurate coupled-cluster methods and 
to calculation of molecular properties.

\begin{acknowledgments}
I would like to thank Prof. B. Jeziorski for fruitful discussions and for reading and commenting 
on the manuscript prior to submission. This work was supported by the National Science Center, 
Poland within the project 2017/27/B/ST4/02739.
\end{acknowledgments}

\nocite{*}
\bibliography{dfcorr}

\begin{thebibliography}{63}%
\makeatletter
\providecommand \@ifxundefined [1]{%
 \@ifx{#1\undefined}
}%
\providecommand \@ifnum [1]{%
 \ifnum #1\expandafter \@firstoftwo
 \else \expandafter \@secondoftwo
 \fi
}%
\providecommand \@ifx [1]{%
 \ifx #1\expandafter \@firstoftwo
 \else \expandafter \@secondoftwo
 \fi
}%
\providecommand \natexlab [1]{#1}%
\providecommand \enquote  [1]{``#1''}%
\providecommand \bibnamefont  [1]{#1}%
\providecommand \bibfnamefont [1]{#1}%
\providecommand \citenamefont [1]{#1}%
\providecommand \href@noop [0]{\@secondoftwo}%
\providecommand \href [0]{\begingroup \@sanitize@url \@href}%
\providecommand \@href[1]{\@@startlink{#1}\@@href}%
\providecommand \@@href[1]{\endgroup#1\@@endlink}%
\providecommand \@sanitize@url [0]{\catcode `\\12\catcode `\$12\catcode
  `\&12\catcode `\#12\catcode `\^12\catcode `\_12\catcode `\%12\relax}%
\providecommand \@@startlink[1]{}%
\providecommand \@@endlink[0]{}%
\providecommand \url  [0]{\begingroup\@sanitize@url \@url }%
\providecommand \@url [1]{\endgroup\@href {#1}{\urlprefix }}%
\providecommand \urlprefix  [0]{URL }%
\providecommand \Eprint [0]{\href }%
\providecommand \doibase [0]{http://dx.doi.org/}%
\providecommand \selectlanguage [0]{\@gobble}%
\providecommand \bibinfo  [0]{\@secondoftwo}%
\providecommand \bibfield  [0]{\@secondoftwo}%
\providecommand \translation [1]{[#1]}%
\providecommand \BibitemOpen [0]{}%
\providecommand \bibitemStop [0]{}%
\providecommand \bibitemNoStop [0]{.\EOS\space}%
\providecommand \EOS [0]{\spacefactor3000\relax}%
\providecommand \BibitemShut  [1]{\csname bibitem#1\endcsname}%
\let\auto@bib@innerbib\@empty
\bibitem [{\citenamefont {Whitten}(1973)}]{whitten73}%
  \BibitemOpen
  \bibfield  {author} {\bibinfo {author} {\bibfnamefont {J.~L.}\ \bibnamefont
  {Whitten}},\ }\href@noop {} {\bibfield  {journal} {\bibinfo  {journal} {J.
  Chem. Phys.}\ }\textbf {\bibinfo {volume} {58}},\ \bibinfo {pages} {4496}
  (\bibinfo {year} {1973})}\BibitemShut {NoStop}%
\bibitem [{\citenamefont {Baerends}, \citenamefont {Ellis},\ and\ \citenamefont
  {Ros}(1973)}]{baerends73}%
  \BibitemOpen
  \bibfield  {author} {\bibinfo {author} {\bibfnamefont {E.}~\bibnamefont
  {Baerends}}, \bibinfo {author} {\bibfnamefont {D.}~\bibnamefont {Ellis}}, \
  and\ \bibinfo {author} {\bibfnamefont {P.}~\bibnamefont {Ros}},\ }\href
  {http://www.sciencedirect.com/science/article/pii/030101047380059X}
  {\bibfield  {journal} {\bibinfo  {journal} {Chem. Phys.}\ }\textbf {\bibinfo
  {volume} {2}},\ \bibinfo {pages} {41 } (\bibinfo {year} {1973})}\BibitemShut
  {NoStop}%
\bibitem [{\citenamefont {Dunlap}, \citenamefont {Connolly},\ and\
  \citenamefont {Sabin}(1979)}]{dunlap79}%
  \BibitemOpen
  \bibfield  {author} {\bibinfo {author} {\bibfnamefont {B.~I.}\ \bibnamefont
  {Dunlap}}, \bibinfo {author} {\bibfnamefont {J.~W.~D.}\ \bibnamefont
  {Connolly}}, \ and\ \bibinfo {author} {\bibfnamefont {J.~R.}\ \bibnamefont
  {Sabin}},\ }\href@noop {} {\bibfield  {journal} {\bibinfo  {journal} {J.
  Chem. Phys.}\ }\textbf {\bibinfo {volume} {71}},\ \bibinfo {pages} {4993}
  (\bibinfo {year} {1979})}\BibitemShut {NoStop}%
\bibitem [{\citenamefont {Van~Alsenoy}(1988)}]{alsenoy88}%
  \BibitemOpen
  \bibfield  {author} {\bibinfo {author} {\bibfnamefont {C.}~\bibnamefont
  {Van~Alsenoy}},\ }\href
  {https://onlinelibrary.wiley.com/doix/abs/10.1002/jcc.540090607} {\bibfield
  {journal} {\bibinfo  {journal} {J. Comp. Chem.}\ }\textbf {\bibinfo {volume}
  {9}},\ \bibinfo {pages} {620} (\bibinfo {year} {1988})}\BibitemShut {NoStop}%
\bibitem [{\citenamefont {Vahtras}, \citenamefont {Almlöf},\ and\
  \citenamefont {Feyereisen}(1993)}]{vahtras93}%
  \BibitemOpen
  \bibfield  {author} {\bibinfo {author} {\bibfnamefont {O.}~\bibnamefont
  {Vahtras}}, \bibinfo {author} {\bibfnamefont {J.}~\bibnamefont {Almlöf}}, \
  and\ \bibinfo {author} {\bibfnamefont {M.}~\bibnamefont {Feyereisen}},\
  }\href {http://www.sciencedirect.com/science/article/pii/0009261493891517}
  {\bibfield  {journal} {\bibinfo  {journal} {Chem. Phys. Lett.}\ }\textbf
  {\bibinfo {volume} {213}},\ \bibinfo {pages} {514 } (\bibinfo {year}
  {1993})}\BibitemShut {NoStop}%
\bibitem [{\citenamefont {Katouda}\ and\ \citenamefont
  {Nagase}(2009)}]{katouda09}%
  \BibitemOpen
  \bibfield  {author} {\bibinfo {author} {\bibfnamefont {M.}~\bibnamefont
  {Katouda}}\ and\ \bibinfo {author} {\bibfnamefont {S.}~\bibnamefont
  {Nagase}},\ }\href
  {https://onlinelibrary.wiley.com/doix/abs/10.1002/qua.22068} {\bibfield
  {journal} {\bibinfo  {journal} {Int. J. Quantum Chem.}\ }\textbf {\bibinfo
  {volume} {109}},\ \bibinfo {pages} {2121} (\bibinfo {year}
  {2009})}\BibitemShut {NoStop}%
\bibitem [{\citenamefont {Manby}(2003)}]{manby03}%
  \BibitemOpen
  \bibfield  {author} {\bibinfo {author} {\bibfnamefont {F.~R.}\ \bibnamefont
  {Manby}},\ }\href@noop {} {\bibfield  {journal} {\bibinfo  {journal} {J.
  Chem. Phys.}\ }\textbf {\bibinfo {volume} {119}},\ \bibinfo {pages} {4607}
  (\bibinfo {year} {2003})}\BibitemShut {NoStop}%
\bibitem [{\citenamefont {Werner}, \citenamefont {Adler},\ and\ \citenamefont
  {Manby}(2007)}]{werner07}%
  \BibitemOpen
  \bibfield  {author} {\bibinfo {author} {\bibfnamefont {H.-J.}\ \bibnamefont
  {Werner}}, \bibinfo {author} {\bibfnamefont {T.~B.}\ \bibnamefont {Adler}}, \
  and\ \bibinfo {author} {\bibfnamefont {F.~R.}\ \bibnamefont {Manby}},\
  }\href@noop {} {\bibfield  {journal} {\bibinfo  {journal} {J. Chem. Phys.}\
  }\textbf {\bibinfo {volume} {126}},\ \bibinfo {pages} {164102} (\bibinfo
  {year} {2007})}\BibitemShut {NoStop}%
\bibitem [{\citenamefont {Schütz}\ and\ \citenamefont
  {Manby}(2003)}]{schutz03}%
  \BibitemOpen
  \bibfield  {author} {\bibinfo {author} {\bibfnamefont {M.}~\bibnamefont
  {Schütz}}\ and\ \bibinfo {author} {\bibfnamefont {F.~R.}\ \bibnamefont
  {Manby}},\ }\href {http://dx.doix.org/10.1039/B304550A} {\bibfield  {journal}
  {\bibinfo  {journal} {Phys. Chem. Chem. Phys.}\ }\textbf {\bibinfo {volume}
  {5}},\ \bibinfo {pages} {3349} (\bibinfo {year} {2003})}\BibitemShut
  {NoStop}%
\bibitem [{\citenamefont {Kats}, \citenamefont {Korona},\ and\ \citenamefont
  {Schütz}(2007)}]{kats07}%
  \BibitemOpen
  \bibfield  {author} {\bibinfo {author} {\bibfnamefont {D.}~\bibnamefont
  {Kats}}, \bibinfo {author} {\bibfnamefont {T.}~\bibnamefont {Korona}}, \ and\
  \bibinfo {author} {\bibfnamefont {M.}~\bibnamefont {Schütz}},\ }\href@noop
  {} {\bibfield  {journal} {\bibinfo  {journal} {J. Chem. Phys.}\ }\textbf
  {\bibinfo {volume} {127}},\ \bibinfo {pages} {064107} (\bibinfo {year}
  {2007})}\BibitemShut {NoStop}%
\bibitem [{\citenamefont {Korona}\ and\ \citenamefont
  {Jeziorski}(2008)}]{korona08}%
  \BibitemOpen
  \bibfield  {author} {\bibinfo {author} {\bibfnamefont {T.}~\bibnamefont
  {Korona}}\ and\ \bibinfo {author} {\bibfnamefont {B.}~\bibnamefont
  {Jeziorski}},\ }\href@noop {} {\bibfield  {journal} {\bibinfo  {journal} {J.
  Chem. Phys.}\ }\textbf {\bibinfo {volume} {128}},\ \bibinfo {pages} {144107}
  (\bibinfo {year} {2008})}\BibitemShut {NoStop}%
\bibitem [{\citenamefont {Boström}\ \emph {et~al.}(2012)\citenamefont
  {Boström}, \citenamefont {Pitoňák}, \citenamefont {Aquilante},
  \citenamefont {Neogrády}, \citenamefont {Pedersen},\ and\ \citenamefont
  {Lindh}}]{bostrom12}%
  \BibitemOpen
  \bibfield  {author} {\bibinfo {author} {\bibfnamefont {J.}~\bibnamefont
  {Boström}}, \bibinfo {author} {\bibfnamefont {M.}~\bibnamefont {Pitoňák}},
  \bibinfo {author} {\bibfnamefont {F.}~\bibnamefont {Aquilante}}, \bibinfo
  {author} {\bibfnamefont {P.}~\bibnamefont {Neogrády}}, \bibinfo {author}
  {\bibfnamefont {T.~B.}\ \bibnamefont {Pedersen}}, \ and\ \bibinfo {author}
  {\bibfnamefont {R.}~\bibnamefont {Lindh}},\ }\href@noop {} {\bibfield
  {journal} {\bibinfo  {journal} {J. Chem. Theory Comp.}\ }\textbf {\bibinfo
  {volume} {8}},\ \bibinfo {pages} {1921} (\bibinfo {year} {2012})}\BibitemShut
  {NoStop}%
\bibitem [{\citenamefont {DePrince}\ and\ \citenamefont
  {Sherrill}(2013)}]{deprince13}%
  \BibitemOpen
  \bibfield  {author} {\bibinfo {author} {\bibfnamefont {A.~E.}\ \bibnamefont
  {DePrince}}\ and\ \bibinfo {author} {\bibfnamefont {C.~D.}\ \bibnamefont
  {Sherrill}},\ }\href@noop {} {\bibfield  {journal} {\bibinfo  {journal} {J.
  Chem. Theory Comp.}\ }\textbf {\bibinfo {volume} {9}},\ \bibinfo {pages}
  {2687} (\bibinfo {year} {2013})}\BibitemShut {NoStop}%
\bibitem [{\citenamefont {Epifanovsky}\ \emph {et~al.}(2013)\citenamefont
  {Epifanovsky}, \citenamefont {Zuev}, \citenamefont {Feng}, \citenamefont
  {Khistyaev}, \citenamefont {Shao},\ and\ \citenamefont
  {Krylov}}]{epifanovsky13}%
  \BibitemOpen
  \bibfield  {author} {\bibinfo {author} {\bibfnamefont {E.}~\bibnamefont
  {Epifanovsky}}, \bibinfo {author} {\bibfnamefont {D.}~\bibnamefont {Zuev}},
  \bibinfo {author} {\bibfnamefont {X.}~\bibnamefont {Feng}}, \bibinfo {author}
  {\bibfnamefont {K.}~\bibnamefont {Khistyaev}}, \bibinfo {author}
  {\bibfnamefont {Y.}~\bibnamefont {Shao}}, \ and\ \bibinfo {author}
  {\bibfnamefont {A.~I.}\ \bibnamefont {Krylov}},\ }\href@noop {} {\bibfield
  {journal} {\bibinfo  {journal} {J. Chem. Phys.}\ }\textbf {\bibinfo {volume}
  {139}},\ \bibinfo {pages} {134105} (\bibinfo {year} {2013})}\BibitemShut
  {NoStop}%
\bibitem [{\citenamefont {A.~Eugene~DePrince}\ \emph
  {et~al.}(2014)\citenamefont {A.~Eugene~DePrince}, \citenamefont {Kennedy},
  \citenamefont {Sumpter},\ and\ \citenamefont {Sherrill}}]{deprince14}%
  \BibitemOpen
  \bibfield  {author} {\bibinfo {author} {\bibfnamefont {I.}~\bibnamefont
  {A.~Eugene~DePrince}}, \bibinfo {author} {\bibfnamefont {M.~R.}\ \bibnamefont
  {Kennedy}}, \bibinfo {author} {\bibfnamefont {B.~G.}\ \bibnamefont
  {Sumpter}}, \ and\ \bibinfo {author} {\bibfnamefont {C.~D.}\ \bibnamefont
  {Sherrill}},\ }\href@noop {} {\bibfield  {journal} {\bibinfo  {journal} {Mol.
  Phys.}\ }\textbf {\bibinfo {volume} {112}},\ \bibinfo {pages} {844} (\bibinfo
  {year} {2014})}\BibitemShut {NoStop}%
\bibitem [{\citenamefont {Weigend}, \citenamefont {Kattannek},\ and\
  \citenamefont {Ahlrichs}(2009)}]{wiegend09}%
  \BibitemOpen
  \bibfield  {author} {\bibinfo {author} {\bibfnamefont {F.}~\bibnamefont
  {Weigend}}, \bibinfo {author} {\bibfnamefont {M.}~\bibnamefont {Kattannek}},
  \ and\ \bibinfo {author} {\bibfnamefont {R.}~\bibnamefont {Ahlrichs}},\
  }\href@noop {} {\bibfield  {journal} {\bibinfo  {journal} {J. Chem. Phys.}\
  }\textbf {\bibinfo {volume} {130}},\ \bibinfo {pages} {164106} (\bibinfo
  {year} {2009})}\BibitemShut {NoStop}%
\bibitem [{\citenamefont {Grajciar}(2015)}]{grajciar15}%
  \BibitemOpen
  \bibfield  {author} {\bibinfo {author} {\bibfnamefont {L.}~\bibnamefont
  {Grajciar}},\ }\href
  {https://onlinelibrary.wiley.com/doix/abs/10.1002/jcc.23961} {\bibfield
  {journal} {\bibinfo  {journal} {J. Comp. Chem.}\ }\textbf {\bibinfo {volume}
  {36}},\ \bibinfo {pages} {1521} (\bibinfo {year} {2015})}\BibitemShut
  {NoStop}%
\bibitem [{\citenamefont {Manzer}\ \emph {et~al.}(2015)\citenamefont {Manzer},
  \citenamefont {Horn}, \citenamefont {Mardirossian},\ and\ \citenamefont
  {Head-Gordon}}]{manzer15}%
  \BibitemOpen
  \bibfield  {author} {\bibinfo {author} {\bibfnamefont {S.}~\bibnamefont
  {Manzer}}, \bibinfo {author} {\bibfnamefont {P.~R.}\ \bibnamefont {Horn}},
  \bibinfo {author} {\bibfnamefont {N.}~\bibnamefont {Mardirossian}}, \ and\
  \bibinfo {author} {\bibfnamefont {M.}~\bibnamefont {Head-Gordon}},\
  }\href@noop {} {\bibfield  {journal} {\bibinfo  {journal} {J. Chem. Phys.}\
  }\textbf {\bibinfo {volume} {143}},\ \bibinfo {pages} {024113} (\bibinfo
  {year} {2015})}\BibitemShut {NoStop}%
\bibitem [{\citenamefont {Duchemin}, \citenamefont {Li},\ and\ \citenamefont
  {Blase}(2017)}]{duchemin17}%
  \BibitemOpen
  \bibfield  {author} {\bibinfo {author} {\bibfnamefont {I.}~\bibnamefont
  {Duchemin}}, \bibinfo {author} {\bibfnamefont {J.}~\bibnamefont {Li}}, \ and\
  \bibinfo {author} {\bibfnamefont {X.}~\bibnamefont {Blase}},\ }\href@noop {}
  {\bibfield  {journal} {\bibinfo  {journal} {J. Chem. Theory Comp.}\ }\textbf
  {\bibinfo {volume} {13}},\ \bibinfo {pages} {1199} (\bibinfo {year}
  {2017})}\BibitemShut {NoStop}%
\bibitem [{\citenamefont {Hohenstein}, \citenamefont {Parrish},\ and\
  \citenamefont {Martínez}(2012)}]{hohenstein12a}%
  \BibitemOpen
  \bibfield  {author} {\bibinfo {author} {\bibfnamefont {E.~G.}\ \bibnamefont
  {Hohenstein}}, \bibinfo {author} {\bibfnamefont {R.~M.}\ \bibnamefont
  {Parrish}}, \ and\ \bibinfo {author} {\bibfnamefont {T.~J.}\ \bibnamefont
  {Martínez}},\ }\href@noop {} {\bibfield  {journal} {\bibinfo  {journal} {J.
  Chem. Phys.}\ }\textbf {\bibinfo {volume} {137}},\ \bibinfo {pages} {044103}
  (\bibinfo {year} {2012})}\BibitemShut {NoStop}%
\bibitem [{\citenamefont {Parrish}\ \emph {et~al.}(2012)\citenamefont
  {Parrish}, \citenamefont {Hohenstein}, \citenamefont {Mart\'{\i}nez},\ and\
  \citenamefont {Sherrill}}]{parrish12}%
  \BibitemOpen
  \bibfield  {author} {\bibinfo {author} {\bibfnamefont {R.~M.}\ \bibnamefont
  {Parrish}}, \bibinfo {author} {\bibfnamefont {E.~G.}\ \bibnamefont
  {Hohenstein}}, \bibinfo {author} {\bibfnamefont {T.~J.}\ \bibnamefont
  {Mart\'{\i}nez}}, \ and\ \bibinfo {author} {\bibfnamefont {C.~D.}\
  \bibnamefont {Sherrill}},\ }\href@noop {} {\bibfield  {journal} {\bibinfo
  {journal} {J. Chem. Phys.}\ }\textbf {\bibinfo {volume} {137}},\ \bibinfo
  {pages} {224106} (\bibinfo {year} {2012})}\BibitemShut {NoStop}%
\bibitem [{\citenamefont {Hohenstein}\ \emph {et~al.}(2012)\citenamefont
  {Hohenstein}, \citenamefont {Parrish}, \citenamefont {Sherrill},\ and\
  \citenamefont {Martínez}}]{hohenstein12b}%
  \BibitemOpen
  \bibfield  {author} {\bibinfo {author} {\bibfnamefont {E.~G.}\ \bibnamefont
  {Hohenstein}}, \bibinfo {author} {\bibfnamefont {R.~M.}\ \bibnamefont
  {Parrish}}, \bibinfo {author} {\bibfnamefont {C.~D.}\ \bibnamefont
  {Sherrill}}, \ and\ \bibinfo {author} {\bibfnamefont {T.~J.}\ \bibnamefont
  {Martínez}},\ }\href@noop {} {\bibfield  {journal} {\bibinfo  {journal} {J.
  Chem. Phys.}\ }\textbf {\bibinfo {volume} {137}},\ \bibinfo {pages} {221101}
  (\bibinfo {year} {2012})}\BibitemShut {NoStop}%
\bibitem [{\citenamefont {Weigend}, \citenamefont {Köhn},\ and\ \citenamefont
  {Hättig}(2002)}]{weigend02}%
  \BibitemOpen
  \bibfield  {author} {\bibinfo {author} {\bibfnamefont {F.}~\bibnamefont
  {Weigend}}, \bibinfo {author} {\bibfnamefont {A.}~\bibnamefont {Köhn}}, \
  and\ \bibinfo {author} {\bibfnamefont {C.}~\bibnamefont {Hättig}},\
  }\href@noop {} {\bibfield  {journal} {\bibinfo  {journal} {J. Chem. Phys.}\
  }\textbf {\bibinfo {volume} {116}},\ \bibinfo {pages} {3175} (\bibinfo {year}
  {2002})}\BibitemShut {NoStop}%
\bibitem [{\citenamefont {Jensen}(2005)}]{jensen05}%
  \BibitemOpen
  \bibfield  {author} {\bibinfo {author} {\bibfnamefont {F.}~\bibnamefont
  {Jensen}},\ }\href {https://doix.org/10.1007/s00214-005-0635-2} {\bibfield
  {journal} {\bibinfo  {journal} {Theor. Chem. Acc.}\ }\textbf {\bibinfo
  {volume} {113}},\ \bibinfo {pages} {267} (\bibinfo {year}
  {2005})}\BibitemShut {NoStop}%
\bibitem [{\citenamefont {H{\"{a}}ttig}(2005)}]{hattig05}%
  \BibitemOpen
  \bibfield  {author} {\bibinfo {author} {\bibfnamefont {C.}~\bibnamefont
  {H{\"{a}}ttig}},\ }\href {http://dx.doix.org/10.1039/B415208E} {\bibfield
  {journal} {\bibinfo  {journal} {Phys. Chem. Chem. Phys.}\ }\textbf {\bibinfo
  {volume} {7}},\ \bibinfo {pages} {59} (\bibinfo {year} {2005})}\BibitemShut
  {NoStop}%
\bibitem [{\citenamefont {Stoychev}, \citenamefont {Auer},\ and\ \citenamefont
  {Neese}(2017)}]{stoychev17}%
  \BibitemOpen
  \bibfield  {author} {\bibinfo {author} {\bibfnamefont {G.~L.}\ \bibnamefont
  {Stoychev}}, \bibinfo {author} {\bibfnamefont {A.~A.}\ \bibnamefont {Auer}},
  \ and\ \bibinfo {author} {\bibfnamefont {F.}~\bibnamefont {Neese}},\
  }\href@noop {} {\bibfield  {journal} {\bibinfo  {journal} {J. Chem. Theory
  Comp.}\ }\textbf {\bibinfo {volume} {13}},\ \bibinfo {pages} {554} (\bibinfo
  {year} {2017})}\BibitemShut {NoStop}%
\bibitem [{\citenamefont {Schurkus}, \citenamefont {Luenser},\ and\
  \citenamefont {Ochsenfeld}(2017)}]{schurkus17}%
  \BibitemOpen
  \bibfield  {author} {\bibinfo {author} {\bibfnamefont {H.~F.}\ \bibnamefont
  {Schurkus}}, \bibinfo {author} {\bibfnamefont {A.}~\bibnamefont {Luenser}}, \
  and\ \bibinfo {author} {\bibfnamefont {C.}~\bibnamefont {Ochsenfeld}},\
  }\href@noop {} {\bibfield  {journal} {\bibinfo  {journal} {J. Chem. Phys.}\
  }\textbf {\bibinfo {volume} {146}},\ \bibinfo {pages} {211106} (\bibinfo
  {year} {2017})}\BibitemShut {NoStop}%
\bibitem [{\citenamefont {Beebe}\ and\ \citenamefont
  {Linderberg}(1977)}]{beebe77}%
  \BibitemOpen
  \bibfield  {author} {\bibinfo {author} {\bibfnamefont {N.~H.~F.}\
  \bibnamefont {Beebe}}\ and\ \bibinfo {author} {\bibfnamefont
  {J.}~\bibnamefont {Linderberg}},\ }\href
  {https://onlinelibrary.wiley.com/doix/abs/10.1002/qua.560120408} {\bibfield
  {journal} {\bibinfo  {journal} {Int. J. Quantum Chem.}\ }\textbf {\bibinfo
  {volume} {12}},\ \bibinfo {pages} {683} (\bibinfo {year} {1977})}\BibitemShut
  {NoStop}%
\bibitem [{\citenamefont {Koch}, \citenamefont {Sánchez~de Merás},\ and\
  \citenamefont {Pedersen}(2003)}]{koch03}%
  \BibitemOpen
  \bibfield  {author} {\bibinfo {author} {\bibfnamefont {H.}~\bibnamefont
  {Koch}}, \bibinfo {author} {\bibfnamefont {A.}~\bibnamefont {Sánchez~de
  Merás}}, \ and\ \bibinfo {author} {\bibfnamefont {T.~B.}\ \bibnamefont
  {Pedersen}},\ }\href@noop {} {\bibfield  {journal} {\bibinfo  {journal} {J.
  Chem. Phys.}\ }\textbf {\bibinfo {volume} {118}},\ \bibinfo {pages} {9481}
  (\bibinfo {year} {2003})}\BibitemShut {NoStop}%
\bibitem [{\citenamefont {Pedersen}, \citenamefont {Sánchez~de Merás},\ and\
  \citenamefont {Koch}(2004)}]{pedersen04}%
  \BibitemOpen
  \bibfield  {author} {\bibinfo {author} {\bibfnamefont {T.~B.}\ \bibnamefont
  {Pedersen}}, \bibinfo {author} {\bibfnamefont {A.~M.~J.}\ \bibnamefont
  {Sánchez~de Merás}}, \ and\ \bibinfo {author} {\bibfnamefont
  {H.}~\bibnamefont {Koch}},\ }\href@noop {} {\bibfield  {journal} {\bibinfo
  {journal} {J. Chem. Phys.}\ }\textbf {\bibinfo {volume} {120}},\ \bibinfo
  {pages} {8887} (\bibinfo {year} {2004})}\BibitemShut {NoStop}%
\bibitem [{\citenamefont {Folkestad}, \citenamefont {Kjønstad},\ and\
  \citenamefont {Koch}(2019)}]{folkestad19}%
  \BibitemOpen
  \bibfield  {author} {\bibinfo {author} {\bibfnamefont {S.~D.}\ \bibnamefont
  {Folkestad}}, \bibinfo {author} {\bibfnamefont {E.~F.}\ \bibnamefont
  {Kjønstad}}, \ and\ \bibinfo {author} {\bibfnamefont {H.}~\bibnamefont
  {Koch}},\ }\href@noop {} {\bibfield  {journal} {\bibinfo  {journal} {J. Chem.
  Phys.}\ }\textbf {\bibinfo {volume} {150}},\ \bibinfo {pages} {194112}
  (\bibinfo {year} {2019})}\BibitemShut {NoStop}%
\bibitem [{\citenamefont {Friesner}(1986)}]{friesner86}%
  \BibitemOpen
  \bibfield  {author} {\bibinfo {author} {\bibfnamefont {R.~A.}\ \bibnamefont
  {Friesner}},\ }\href@noop {} {\bibfield  {journal} {\bibinfo  {journal} {J.
  Chem. Phys.}\ }\textbf {\bibinfo {volume} {85}},\ \bibinfo {pages} {1462}
  (\bibinfo {year} {1986})}\BibitemShut {NoStop}%
\bibitem [{\citenamefont {Martinez}\ and\ \citenamefont
  {Carter}(1994)}]{martinez94}%
  \BibitemOpen
  \bibfield  {author} {\bibinfo {author} {\bibfnamefont {T.~J.}\ \bibnamefont
  {Martinez}}\ and\ \bibinfo {author} {\bibfnamefont {E.~A.}\ \bibnamefont
  {Carter}},\ }\href@noop {} {\bibfield  {journal} {\bibinfo  {journal} {J.
  Chem. Phys.}\ }\textbf {\bibinfo {volume} {100}},\ \bibinfo {pages} {3631}
  (\bibinfo {year} {1994})}\BibitemShut {NoStop}%
\bibitem [{\citenamefont {Parrish}\ \emph {et~al.}(2013)\citenamefont
  {Parrish}, \citenamefont {Hohenstein}, \citenamefont {Schunck}, \citenamefont
  {Sherrill},\ and\ \citenamefont {Mart\'{\i}nez}}]{parrish13}%
  \BibitemOpen
  \bibfield  {author} {\bibinfo {author} {\bibfnamefont {R.~M.}\ \bibnamefont
  {Parrish}}, \bibinfo {author} {\bibfnamefont {E.~G.}\ \bibnamefont
  {Hohenstein}}, \bibinfo {author} {\bibfnamefont {N.~F.}\ \bibnamefont
  {Schunck}}, \bibinfo {author} {\bibfnamefont {C.~D.}\ \bibnamefont
  {Sherrill}}, \ and\ \bibinfo {author} {\bibfnamefont {T.~J.}\ \bibnamefont
  {Mart\'{\i}nez}},\ }\href
  {https://link.aps.org/doix/10.1103/PhysRevLett.111.132505} {\bibfield
  {journal} {\bibinfo  {journal} {Phys. Rev. Lett.}\ }\textbf {\bibinfo
  {volume} {111}},\ \bibinfo {pages} {132505} (\bibinfo {year}
  {2013})}\BibitemShut {NoStop}%
\bibitem [{\citenamefont {Benedikt}\ \emph {et~al.}(2011)\citenamefont
  {Benedikt}, \citenamefont {Auer}, \citenamefont {Espig},\ and\ \citenamefont
  {Hackbusch}}]{benedikt11}%
  \BibitemOpen
  \bibfield  {author} {\bibinfo {author} {\bibfnamefont {U.}~\bibnamefont
  {Benedikt}}, \bibinfo {author} {\bibfnamefont {A.~A.}\ \bibnamefont {Auer}},
  \bibinfo {author} {\bibfnamefont {M.}~\bibnamefont {Espig}}, \ and\ \bibinfo
  {author} {\bibfnamefont {W.}~\bibnamefont {Hackbusch}},\ }\href@noop {}
  {\bibfield  {journal} {\bibinfo  {journal} {J. Chem. Phys.}\ }\textbf
  {\bibinfo {volume} {134}},\ \bibinfo {pages} {054118} (\bibinfo {year}
  {2011})}\BibitemShut {NoStop}%
\bibitem [{\citenamefont {Benedikt}, \citenamefont {B\"{o}hm},\ and\
  \citenamefont {Auer}(2013)}]{benedikt13}%
  \BibitemOpen
  \bibfield  {author} {\bibinfo {author} {\bibfnamefont {U.}~\bibnamefont
  {Benedikt}}, \bibinfo {author} {\bibfnamefont {K.-H.}\ \bibnamefont
  {B\"{o}hm}}, \ and\ \bibinfo {author} {\bibfnamefont {A.~A.}\ \bibnamefont
  {Auer}},\ }\href@noop {} {\bibfield  {journal} {\bibinfo  {journal} {J. Chem.
  Phys.}\ }\textbf {\bibinfo {volume} {139}},\ \bibinfo {pages} {224101}
  (\bibinfo {year} {2013})}\BibitemShut {NoStop}%
\bibitem [{\citenamefont {Neese}\ \emph {et~al.}(2009)\citenamefont {Neese},
  \citenamefont {Wennmohs}, \citenamefont {Hansen},\ and\ \citenamefont
  {Becker}}]{neese09}%
  \BibitemOpen
  \bibfield  {author} {\bibinfo {author} {\bibfnamefont {F.}~\bibnamefont
  {Neese}}, \bibinfo {author} {\bibfnamefont {F.}~\bibnamefont {Wennmohs}},
  \bibinfo {author} {\bibfnamefont {A.}~\bibnamefont {Hansen}}, \ and\ \bibinfo
  {author} {\bibfnamefont {U.}~\bibnamefont {Becker}},\ }\href
  {http://www.sciencedirect.com/science/article/pii/S0301010408005089}
  {\bibfield  {journal} {\bibinfo  {journal} {Chem. Phys.}\ }\textbf {\bibinfo
  {volume} {356}},\ \bibinfo {pages} {98 } (\bibinfo {year}
  {2009})}\BibitemShut {NoStop}%
\bibitem [{\citenamefont {Kossmann}\ and\ \citenamefont
  {Neese}(2010)}]{kossmann10}%
  \BibitemOpen
  \bibfield  {author} {\bibinfo {author} {\bibfnamefont {S.}~\bibnamefont
  {Kossmann}}\ and\ \bibinfo {author} {\bibfnamefont {F.}~\bibnamefont
  {Neese}},\ }\href@noop {} {\bibfield  {journal} {\bibinfo  {journal} {J.
  Chem. Theory Comp.}\ }\textbf {\bibinfo {volume} {6}},\ \bibinfo {pages}
  {2325} (\bibinfo {year} {2010})}\BibitemShut {NoStop}%
\bibitem [{\citenamefont {Izs\'{a}k}\ and\ \citenamefont
  {Neese}(2011)}]{izsak11}%
  \BibitemOpen
  \bibfield  {author} {\bibinfo {author} {\bibfnamefont {R.}~\bibnamefont
  {Izs\'{a}k}}\ and\ \bibinfo {author} {\bibfnamefont {F.}~\bibnamefont
  {Neese}},\ }\href@noop {} {\bibfield  {journal} {\bibinfo  {journal} {J.
  Chem. Phys.}\ }\textbf {\bibinfo {volume} {135}},\ \bibinfo {pages} {144105}
  (\bibinfo {year} {2011})}\BibitemShut {NoStop}%
\bibitem [{\citenamefont {Petrenko}, \citenamefont {Kossmann},\ and\
  \citenamefont {Neese}(2011)}]{taras11}%
  \BibitemOpen
  \bibfield  {author} {\bibinfo {author} {\bibfnamefont {T.}~\bibnamefont
  {Petrenko}}, \bibinfo {author} {\bibfnamefont {S.}~\bibnamefont {Kossmann}},
  \ and\ \bibinfo {author} {\bibfnamefont {F.}~\bibnamefont {Neese}},\
  }\href@noop {} {\bibfield  {journal} {\bibinfo  {journal} {J. Chem. Phys.}\
  }\textbf {\bibinfo {volume} {134}},\ \bibinfo {pages} {054116} (\bibinfo
  {year} {2011})}\BibitemShut {NoStop}%
\bibitem [{\citenamefont {Izs\'{a}k}, \citenamefont {Hansen},\ and\
  \citenamefont {Neese}(2012)}]{izsak12}%
  \BibitemOpen
  \bibfield  {author} {\bibinfo {author} {\bibfnamefont {R.}~\bibnamefont
  {Izs\'{a}k}}, \bibinfo {author} {\bibfnamefont {A.}~\bibnamefont {Hansen}}, \
  and\ \bibinfo {author} {\bibfnamefont {F.}~\bibnamefont {Neese}},\
  }\href@noop {} {\bibfield  {journal} {\bibinfo  {journal} {Mol. Phys.}\
  }\textbf {\bibinfo {volume} {110}},\ \bibinfo {pages} {2413} (\bibinfo {year}
  {2012})}\BibitemShut {NoStop}%
\bibitem [{\citenamefont {Izs\'{a}k}\ and\ \citenamefont
  {Neese}(2013)}]{izsak13}%
  \BibitemOpen
  \bibfield  {author} {\bibinfo {author} {\bibfnamefont {R.}~\bibnamefont
  {Izs\'{a}k}}\ and\ \bibinfo {author} {\bibfnamefont {F.}~\bibnamefont
  {Neese}},\ }\href@noop {} {\bibfield  {journal} {\bibinfo  {journal} {Mol.
  Phys.}\ }\textbf {\bibinfo {volume} {111}},\ \bibinfo {pages} {1190}
  (\bibinfo {year} {2013})}\BibitemShut {NoStop}%
\bibitem [{\citenamefont {Dutta}, \citenamefont {Neese},\ and\ \citenamefont
  {Izs\'{a}k}(2016)}]{dutta16}%
  \BibitemOpen
  \bibfield  {author} {\bibinfo {author} {\bibfnamefont {A.~K.}\ \bibnamefont
  {Dutta}}, \bibinfo {author} {\bibfnamefont {F.}~\bibnamefont {Neese}}, \ and\
  \bibinfo {author} {\bibfnamefont {R.}~\bibnamefont {Izs\'{a}k}},\ }\href@noop
  {} {\bibfield  {journal} {\bibinfo  {journal} {J. Chem. Phys.}\ }\textbf
  {\bibinfo {volume} {144}},\ \bibinfo {pages} {034102} (\bibinfo {year}
  {2016})}\BibitemShut {NoStop}%
\bibitem [{\citenamefont {Crawford}\ and\ \citenamefont
  {Schaefer~III}(2007)}]{crawford07}%
  \BibitemOpen
  \bibfield  {author} {\bibinfo {author} {\bibfnamefont {T.~D.}\ \bibnamefont
  {Crawford}}\ and\ \bibinfo {author} {\bibfnamefont {H.~F.}\ \bibnamefont
  {Schaefer~III}},\ }\enquote {\bibinfo {title} {An introduction to coupled
  cluster theory for computational chemists},}\ in\ \href
  {https://onlinelibrary.wiley.com/doix/abs/10.1002/9780470125915.ch2} {\emph
  {\bibinfo {booktitle} {Reviews in Computational Chemistry}}}\ (\bibinfo
  {publisher} {John Wiley \& Sons, Ltd},\ \bibinfo {year} {2007})\ pp.\
  \bibinfo {pages} {33--136}\BibitemShut {NoStop}%
\bibitem [{\citenamefont {Bartlett}\ and\ \citenamefont
  {Musia\l{}}(2007)}]{bartlett07}%
  \BibitemOpen
  \bibfield  {author} {\bibinfo {author} {\bibfnamefont {R.~J.}\ \bibnamefont
  {Bartlett}}\ and\ \bibinfo {author} {\bibfnamefont {M.}~\bibnamefont
  {Musia\l{}}},\ }\href {https://link.aps.org/doix/10.1103/RevModPhys.79.291}
  {\bibfield  {journal} {\bibinfo  {journal} {Rev. Mod. Phys.}\ }\textbf
  {\bibinfo {volume} {79}},\ \bibinfo {pages} {291} (\bibinfo {year}
  {2007})}\BibitemShut {NoStop}%
\bibitem [{\citenamefont {Purvis}\ and\ \citenamefont
  {Bartlett}(1982)}]{purvis82}%
  \BibitemOpen
  \bibfield  {author} {\bibinfo {author} {\bibfnamefont {G.~D.}\ \bibnamefont
  {Purvis}}\ and\ \bibinfo {author} {\bibfnamefont {R.~J.}\ \bibnamefont
  {Bartlett}},\ }\href@noop {} {\bibfield  {journal} {\bibinfo  {journal} {J.
  Chem. Phys.}\ }\textbf {\bibinfo {volume} {76}},\ \bibinfo {pages} {1910}
  (\bibinfo {year} {1982})}\BibitemShut {NoStop}%
\bibitem [{\citenamefont {Scuseria}\ \emph {et~al.}(1987)\citenamefont
  {Scuseria}, \citenamefont {Scheiner}, \citenamefont {Lee}, \citenamefont
  {Rice},\ and\ \citenamefont {Schaefer}}]{scuseria87}%
  \BibitemOpen
  \bibfield  {author} {\bibinfo {author} {\bibfnamefont {G.~E.}\ \bibnamefont
  {Scuseria}}, \bibinfo {author} {\bibfnamefont {A.~C.}\ \bibnamefont
  {Scheiner}}, \bibinfo {author} {\bibfnamefont {T.~J.}\ \bibnamefont {Lee}},
  \bibinfo {author} {\bibfnamefont {J.~E.}\ \bibnamefont {Rice}}, \ and\
  \bibinfo {author} {\bibfnamefont {H.~F.}\ \bibnamefont {Schaefer}},\
  }\href@noop {} {\bibfield  {journal} {\bibinfo  {journal} {J. Chem. Phys.}\
  }\textbf {\bibinfo {volume} {86}},\ \bibinfo {pages} {2881} (\bibinfo {year}
  {1987})}\BibitemShut {NoStop}%
\bibitem [{\citenamefont {Pedersen}, \citenamefont {Aquilante},\ and\
  \citenamefont {Lindh}(2009)}]{pedersen09}%
  \BibitemOpen
  \bibfield  {author} {\bibinfo {author} {\bibfnamefont {T.~B.}\ \bibnamefont
  {Pedersen}}, \bibinfo {author} {\bibfnamefont {F.}~\bibnamefont {Aquilante}},
  \ and\ \bibinfo {author} {\bibfnamefont {R.}~\bibnamefont {Lindh}},\
  }\href@noop {} {\bibfield  {journal} {\bibinfo  {journal} {Theoretical
  Chemistry Accounts}\ }\textbf {\bibinfo {volume} {124}},\ \bibinfo {pages}
  {1} (\bibinfo {year} {2009})}\BibitemShut {NoStop}%
\bibitem [{\citenamefont {Monkhorst}(1977)}]{monkhorst77}%
  \BibitemOpen
  \bibfield  {author} {\bibinfo {author} {\bibfnamefont {H.~J.}\ \bibnamefont
  {Monkhorst}},\ }\href@noop {} {\bibfield  {journal} {\bibinfo  {journal}
  {Int. J. Quantum Chem.}\ }\textbf {\bibinfo {volume} {12}},\ \bibinfo {pages}
  {421} (\bibinfo {year} {1977})}\BibitemShut {NoStop}%
\bibitem [{\citenamefont {Dalgaard}\ and\ \citenamefont
  {Monkhorst}(1983)}]{dalgaard83}%
  \BibitemOpen
  \bibfield  {author} {\bibinfo {author} {\bibfnamefont {E.}~\bibnamefont
  {Dalgaard}}\ and\ \bibinfo {author} {\bibfnamefont {H.~J.}\ \bibnamefont
  {Monkhorst}},\ }\href {https://link.aps.org/doix/10.1103/PhysRevA.28.1217}
  {\bibfield  {journal} {\bibinfo  {journal} {Phys. Rev. A}\ }\textbf {\bibinfo
  {volume} {28}},\ \bibinfo {pages} {1217} (\bibinfo {year}
  {1983})}\BibitemShut {NoStop}%
\bibitem [{\citenamefont {Koch}\ and\ \citenamefont
  {Jorgensen}(1990)}]{koch90}%
  \BibitemOpen
  \bibfield  {author} {\bibinfo {author} {\bibfnamefont {H.}~\bibnamefont
  {Koch}}\ and\ \bibinfo {author} {\bibfnamefont {P.}~\bibnamefont
  {Jorgensen}},\ }\href@noop {} {\bibfield  {journal} {\bibinfo  {journal} {J.
  Chem. Phys.}\ }\textbf {\bibinfo {volume} {93}},\ \bibinfo {pages} {3333}
  (\bibinfo {year} {1990})}\BibitemShut {NoStop}%
\bibitem [{\citenamefont {Kobayashi}, \citenamefont {Koch},\ and\ \citenamefont
  {Jorgensen}(1994)}]{kobayashi94}%
  \BibitemOpen
  \bibfield  {author} {\bibinfo {author} {\bibfnamefont {R.}~\bibnamefont
  {Kobayashi}}, \bibinfo {author} {\bibfnamefont {H.}~\bibnamefont {Koch}}, \
  and\ \bibinfo {author} {\bibfnamefont {P.}~\bibnamefont {Jorgensen}},\ }\href
  {http://www.sciencedirect.com/science/article/pii/0009261494000514}
  {\bibfield  {journal} {\bibinfo  {journal} {Chem. Phys. Lett.}\ }\textbf
  {\bibinfo {volume} {219}},\ \bibinfo {pages} {30 } (\bibinfo {year}
  {1994})}\BibitemShut {NoStop}%
\bibitem [{\citenamefont {Koch}\ \emph {et~al.}(1994)\citenamefont {Koch},
  \citenamefont {Christiansen}, \citenamefont {Kobayashi}, \citenamefont
  {Jørgensen},\ and\ \citenamefont {Helgaker}}]{koch94}%
  \BibitemOpen
  \bibfield  {author} {\bibinfo {author} {\bibfnamefont {H.}~\bibnamefont
  {Koch}}, \bibinfo {author} {\bibfnamefont {O.}~\bibnamefont {Christiansen}},
  \bibinfo {author} {\bibfnamefont {R.}~\bibnamefont {Kobayashi}}, \bibinfo
  {author} {\bibfnamefont {P.}~\bibnamefont {Jørgensen}}, \ and\ \bibinfo
  {author} {\bibfnamefont {T.}~\bibnamefont {Helgaker}},\ }\href
  {http://www.sciencedirect.com/science/article/pii/0009261494008981}
  {\bibfield  {journal} {\bibinfo  {journal} {Chem. Phys. Lett.}\ }\textbf
  {\bibinfo {volume} {228}},\ \bibinfo {pages} {233 } (\bibinfo {year}
  {1994})}\BibitemShut {NoStop}%
\bibitem [{\citenamefont {Koch}\ \emph {et~al.}(1996)\citenamefont {Koch},
  \citenamefont {Sánchez~de Merás}, \citenamefont {Helgaker},\ and\
  \citenamefont {Christiansen}}]{koch96}%
  \BibitemOpen
  \bibfield  {author} {\bibinfo {author} {\bibfnamefont {H.}~\bibnamefont
  {Koch}}, \bibinfo {author} {\bibfnamefont {A.}~\bibnamefont {Sánchez~de
  Merás}}, \bibinfo {author} {\bibfnamefont {T.}~\bibnamefont {Helgaker}}, \
  and\ \bibinfo {author} {\bibfnamefont {O.}~\bibnamefont {Christiansen}},\
  }\href@noop {} {\bibfield  {journal} {\bibinfo  {journal} {J. Chem. Phys.}\
  }\textbf {\bibinfo {volume} {104}},\ \bibinfo {pages} {4157} (\bibinfo {year}
  {1996})}\BibitemShut {NoStop}%
\bibitem [{\citenamefont {Schmidt}\ \emph {et~al.}(1993)\citenamefont
  {Schmidt}, \citenamefont {Baldridge}, \citenamefont {Boatz}, \citenamefont
  {Elbert}, \citenamefont {Gordon}, \citenamefont {Jensen}, \citenamefont
  {Koseki}, \citenamefont {Matsunaga}, \citenamefont {Nguyen}, \citenamefont
  {Su}, \citenamefont {Windus}, \citenamefont {Dupuis},\ and\ \citenamefont
  {Montgomery~Jr}}]{schmidt93}%
  \BibitemOpen
  \bibfield  {author} {\bibinfo {author} {\bibfnamefont {M.~W.}\ \bibnamefont
  {Schmidt}}, \bibinfo {author} {\bibfnamefont {K.~K.}\ \bibnamefont
  {Baldridge}}, \bibinfo {author} {\bibfnamefont {J.~A.}\ \bibnamefont
  {Boatz}}, \bibinfo {author} {\bibfnamefont {S.~T.}\ \bibnamefont {Elbert}},
  \bibinfo {author} {\bibfnamefont {M.~S.}\ \bibnamefont {Gordon}}, \bibinfo
  {author} {\bibfnamefont {J.~H.}\ \bibnamefont {Jensen}}, \bibinfo {author}
  {\bibfnamefont {S.}~\bibnamefont {Koseki}}, \bibinfo {author} {\bibfnamefont
  {N.}~\bibnamefont {Matsunaga}}, \bibinfo {author} {\bibfnamefont {K.~A.}\
  \bibnamefont {Nguyen}}, \bibinfo {author} {\bibfnamefont {S.}~\bibnamefont
  {Su}}, \bibinfo {author} {\bibfnamefont {T.~L.}\ \bibnamefont {Windus}},
  \bibinfo {author} {\bibfnamefont {M.}~\bibnamefont {Dupuis}}, \ and\ \bibinfo
  {author} {\bibfnamefont {J.~A.}\ \bibnamefont {Montgomery~Jr}},\ }\href@noop
  {} {\bibfield  {journal} {\bibinfo  {journal} {J. Comp. Chem.}\ }\textbf
  {\bibinfo {volume} {14}},\ \bibinfo {pages} {1347} (\bibinfo {year}
  {1993})}\BibitemShut {NoStop}%
\bibitem [{\citenamefont {Dunning}(1989)}]{dunning89}%
  \BibitemOpen
  \bibfield  {author} {\bibinfo {author} {\bibfnamefont {T.~H.}\ \bibnamefont
  {Dunning}},\ }\href@noop {} {\bibfield  {journal} {\bibinfo  {journal} {J.
  Chem. Phys.}\ }\textbf {\bibinfo {volume} {90}},\ \bibinfo {pages} {1007}
  (\bibinfo {year} {1989})}\BibitemShut {NoStop}%
\bibitem [{\citenamefont {Kendall}, \citenamefont {Dunning},\ and\
  \citenamefont {Harrison}(1992)}]{kendall92}%
  \BibitemOpen
  \bibfield  {author} {\bibinfo {author} {\bibfnamefont {R.~A.}\ \bibnamefont
  {Kendall}}, \bibinfo {author} {\bibfnamefont {T.~H.}\ \bibnamefont
  {Dunning}}, \ and\ \bibinfo {author} {\bibfnamefont {R.~J.}\ \bibnamefont
  {Harrison}},\ }\href@noop {} {\bibfield  {journal} {\bibinfo  {journal} {J.
  Chem. Phys.}\ }\textbf {\bibinfo {volume} {96}},\ \bibinfo {pages} {6796}
  (\bibinfo {year} {1992})}\BibitemShut {NoStop}%
\bibitem [{\citenamefont {Woon}\ and\ \citenamefont {Dunning}(1993)}]{woon93}%
  \BibitemOpen
  \bibfield  {author} {\bibinfo {author} {\bibfnamefont {D.~E.}\ \bibnamefont
  {Woon}}\ and\ \bibinfo {author} {\bibfnamefont {T.~H.}\ \bibnamefont
  {Dunning}},\ }\href@noop {} {\bibfield  {journal} {\bibinfo  {journal} {J.
  Chem. Phys.}\ }\textbf {\bibinfo {volume} {98}},\ \bibinfo {pages} {1358}
  (\bibinfo {year} {1993})}\BibitemShut {NoStop}%
\bibitem [{\citenamefont {Weigend}\ \emph {et~al.}(1998)\citenamefont
  {Weigend}, \citenamefont {Häser}, \citenamefont {Patzelt},\ and\
  \citenamefont {Ahlrichs}}]{weigend98}%
  \BibitemOpen
  \bibfield  {author} {\bibinfo {author} {\bibfnamefont {F.}~\bibnamefont
  {Weigend}}, \bibinfo {author} {\bibfnamefont {M.}~\bibnamefont {Häser}},
  \bibinfo {author} {\bibfnamefont {H.}~\bibnamefont {Patzelt}}, \ and\
  \bibinfo {author} {\bibfnamefont {R.}~\bibnamefont {Ahlrichs}},\ }\href
  {http://www.sciencedirect.com/science/article/pii/S0009261498008628}
  {\bibfield  {journal} {\bibinfo  {journal} {Chem. Phys. Lett.}\ }\textbf
  {\bibinfo {volume} {294}},\ \bibinfo {pages} {143 } (\bibinfo {year}
  {1998})}\BibitemShut {NoStop}%
\bibitem [{\citenamefont {Adler}\ and\ \citenamefont {Werner}(2011)}]{adler11}%
  \BibitemOpen
  \bibfield  {author} {\bibinfo {author} {\bibfnamefont {T.~B.}\ \bibnamefont
  {Adler}}\ and\ \bibinfo {author} {\bibfnamefont {H.-J.}\ \bibnamefont
  {Werner}},\ }\href@noop {} {\bibfield  {journal} {\bibinfo  {journal} {J.
  Chem. Phys.}\ }\textbf {\bibinfo {volume} {135}},\ \bibinfo {pages} {144117}
  (\bibinfo {year} {2011})}\BibitemShut {NoStop}%
\bibitem [{\citenamefont {Řezáč}\ and\ \citenamefont
  {Hobza}(2013)}]{rezac13}%
  \BibitemOpen
  \bibfield  {author} {\bibinfo {author} {\bibfnamefont {J.}~\bibnamefont
  {Řezáč}}\ and\ \bibinfo {author} {\bibfnamefont {P.}~\bibnamefont
  {Hobza}},\ }\href@noop {} {\bibfield  {journal} {\bibinfo  {journal} {J.
  Chem. Theory Comp.}\ }\textbf {\bibinfo {volume} {9}},\ \bibinfo {pages}
  {2151} (\bibinfo {year} {2013})}\BibitemShut {NoStop}%
\bibitem [{\citenamefont {Raghavachari}\ \emph {et~al.}(1989)\citenamefont
  {Raghavachari}, \citenamefont {Trucks}, \citenamefont {Pople},\ and\
  \citenamefont {Head-Gordon}}]{ragha89}%
  \BibitemOpen
  \bibfield  {author} {\bibinfo {author} {\bibfnamefont {K.}~\bibnamefont
  {Raghavachari}}, \bibinfo {author} {\bibfnamefont {G.~W.}\ \bibnamefont
  {Trucks}}, \bibinfo {author} {\bibfnamefont {J.~A.}\ \bibnamefont {Pople}}, \
  and\ \bibinfo {author} {\bibfnamefont {M.}~\bibnamefont {Head-Gordon}},\
  }\href {http://www.sciencedirect.com/science/article/pii/S0009261489873956}
  {\bibfield  {journal} {\bibinfo  {journal} {Chem. Phys. Lett.}\ }\textbf
  {\bibinfo {volume} {157}},\ \bibinfo {pages} {479 } (\bibinfo {year}
  {1989})}\BibitemShut {NoStop}%
\bibitem [{\citenamefont {Jurečka}\ \emph {et~al.}(2006)\citenamefont
  {Jurečka}, \citenamefont {Šponer}, \citenamefont {Černý},\ and\
  \citenamefont {Hobza}}]{jurecka06}%
  \BibitemOpen
  \bibfield  {author} {\bibinfo {author} {\bibfnamefont {P.}~\bibnamefont
  {Jurečka}}, \bibinfo {author} {\bibfnamefont {J.}~\bibnamefont {Šponer}},
  \bibinfo {author} {\bibfnamefont {J.}~\bibnamefont {Černý}}, \ and\
  \bibinfo {author} {\bibfnamefont {P.}~\bibnamefont {Hobza}},\ }\href@noop {}
  {\bibfield  {journal} {\bibinfo  {journal} {Phys. Chem. Chem. Phys.}\
  }\textbf {\bibinfo {volume} {8}},\ \bibinfo {pages} {1985} (\bibinfo {year}
  {2006})}\BibitemShut {NoStop}%
\end{thebibliography}%

\end{document}